# Analysis of Neighbourhoods in Multi-layered Dynamic Social Networks

**Piotr Bródka**
*Wrocław University of Technology, Wyb. Wyspiańskiego 27*
*50-370 Wrocław, Poland*
*piotr.brodka@pwr.wroc.pl*

**Przemysław Kazienko**
*Wrocław University of Technology, Wyb. Wyspiańskiego 27*
*50-370 Wrocław, Poland*
*kazienko@pwr.wroc.pl*

**Katarzyna Musial**
*School of Natural and Mathematical Sciences, Department of Informatics, King's College London,*
*London, United Kingdom*
*katarzyna.musial@kcl.ac.uk*

**Krzysztof Skibicki**
*Wrocław University of Technology, Wyb. Wyspiańskiego 27*
*50-370 Wrocław, Poland*
*krzysztof.skibicki@gmail.com)*

## Abstract

Social networks existing among employees, customers or other types of users of various IT systems have become one of the research areas of growing importance. Data about people and their interactions that exist in social media, provides information about many different types of relationships within one network. Analysing this data one can obtain knowledge not only about the structure and characteristics of the network but it also enables to understand the semantic of human relations.

Each social network consists of nodes – social entities and edges linking pairs of nodes. In regular, one-layered networks, two nodes – i.e. people are connected with a single edge whereas in the multi-layered social networks, there may be many links of different types for a pair of nodes. Most of the methods used for social network analysis (SNA) may be applied only to one-layered networks. Thus, some new structural measures for multi-layered social networks are proposed in the paper. This study focuses on definitions and analysis of cross-layer clustering coefficient, cross-layer degree centrality and various versions of multi-layered degree centralities. Authors also investigated the dynamics of multi-layered neighbourhood. The evaluation of the presented concepts on the real-world dataset is presented. The measures proposed in the paper may directly be used to various methods for collective classification, in which nodes are assigned to labels according to their structural input features.

## 1. Introduction

Social network (SN) is an example of complex networked system in which set of nodes (social entities) interact with each other[1,2]. Based on these interactions, their types and intensity, the relationships between nodes can be defined.

The concept of SN is not new and has been researched by scientists from many fields, such as sociology, geography, social psychology, management, computer science and even biology and physics. First studies on social networks can be dated back to the beginning of twentieth century. These analyses focused on small samples (up to few thousands of users) because, due to the lack of computational resources, data had to be manually handled. Also the process of data gathering was very time consuming as it was based on surveys. The increasing amount of available computational power has enabled to analyse more and more data but still researchers spent a lot of time on data collecting process. Nowadays, with the development of a large number of social media, the amount of data that is stored on the computer servers is enormous. All people activities are logged and can be analysed without conducting any additional surveys. This explosion of the amount of available data yields for more and more sophisticated techniques that have to be developed in order to extract the meaningful information.

Data about people activities and interactions, collected in different systems, enable to extract social networks in which different types of relations exist. All these relations should be analysed in parallel as the knowledge is not only hidden in individual layers. The information about what happens on one layer can influence the actions on other layers. This paper is an attempt to analyse this kind of networks as a whole, without neglecting information about different relations types between people. Thus, new measures for investigating multi-layered character of the network are proposed. These are: cross-layer clustering coefficient, cross-layer degree centrality and various versions of multi-layered degree centralities.

To enable the systematic representation of users and different types of connections, authors proposed a structure called multi-layered social network (MSN). The profile of MSN consists of many layers, corresponding to different kinds of relationships and user activities.

In addition, nodes and connections within the layers change over time. It means that the dynamics of these networks is their inseparable feature and cannot be neglected during the analysis. Thus, this study also looks at the changes that occur in different layers of the network in regards to the proposed measures. For both static and dynamic structural analyses that were performed, some semantic of these analyses is investigated.

The rest of the paper is structured as follows: the next section includes the description of related work in the area of social networks where different types of relations were analysed. In section 3, the concept of the multi-layered social network MSN is presented. Afterwards, in section 4 definitions of different multi-layered neighbourhoods are provided. Several new structural measures for multi-layered social networks, in particular cross-layer clustering coefficient and various versions of multi-layered degree centrality are proposed in section 5 and 6, respectively. Experimental studies are depicted in the following section 7 and the paper in concluded in section 8.

## 2. Related Work

Social network can be defined as the finite set of actors (network nodes) and relationships (network edges) that link these actors. This concept is not new and has been researched by people from different fields for many decades[1,2,3,4]. Also the networks where more than one type of relation exists are not new in the world of science[2] and they were analysed mainly at the small scale[5,6,7]. However, recently the area of large-scale multi-layered network has started attracting more and more attention from researchers from different fields[8,9,10,11]. These networks are also known as multi-relational, multiplex, multi-layered or multivariate networks[11].

Social networks emerging from different types of social media are good examples of multi-relational networks. One reason for such a big interest in this area is the fact that these systems offer large datasets including information about peoples' profiles and activities that can be analysed. Due to the fact that this data reflect users' behaviours in the virtual world, the networks extracted from this data are called online social networks[12], web-based social networks[13], or computer-supported social networks[14].

Bibliographic data[15], blogs[16], photos sharing systems like Flickr[8], e-mail systems[17], telecommunication data[18], social services like Twitter[19] or Facebook[20], video sharing systems like YouTube[21], Wikipedia[22] are the examples of data sources which are used by many researches to analyse the underlying social networks. However, this vast amount of data and especially its multi-relational character are the source of new research challenges connected with processing of this data[23]. Although most of the existing methods work properly for single-layered networks, there is a lack of well-established tools for multi-layered network analysis. Development of new metrics is very important from the perspective of further advances in the web science as the multi-relational networks can be found almost everywhere, they are more expressive in terms of the semantic information and give opportunity to analyse different types of human relationships[10].

Researchers usually try to cope with multi-layered large-scale networks by analysing layers separately using the existing methods for one-layered networks and then comparing the results using some correlation measure (e.g. Jaccard coefficient or cosine measure). In Ref. 11 authors distinguished 6 different relation types between users of the massive multiplayer online game. First they analysed the characteristics of each layer separately and after that they studied correlations and overlap between the extracted types of relations. One of the interesting findings is that users tend to play different roles in different networks. From the structural perspective, authors found that different types of interactions are characterised by different patterns of connectivity, e.g. according to their study power-law degree distributions indicate aggressive actions. Another example is the analysis of Flickr[8], where authors distinguished eleven types of relationships between users. First authors investigated layers separately and then used the correlation measures to compare these networks. Their main finding was that depending on the type of the relations they are either semantically or socially driven.

It is also possible to create from multi-layered network a single-layered network and then to apply the existing methods to such structure. An example of such procedure is path algebra that purpose is to transform a multi-relational network to single-relational networks that is "semantically-rich"[10].

Another way to deal with multiplex networks is to develop new methods for their analysis. Such works are usually to some extend based on the existing methods for single-layered networks. The investigated topics are among others: community mining[24,25], ranking network's nodes[26] and paths[27], shortest[28] and unique[29] paths finding.

The researchers in the field of multi-layered networks also try to develop new models of networks that capture the multi-relational characteristics of data. An example can be a multi-layered semantic social network model that enables to investigate human interests in more details than when they are analysed all together[30].

Although, some research has been done in the field of multi-relational social networks, to our best knowledge there has been no work reported on what in this paper is investigated, i.e. multi-layered neighbourhoods. We aim at analysing both static and dynamic characteristics of the individuals' neighbourhoods in multi-layered social networks. Social Network Analysis provides measures to investigate user neighbourhood in a single-layered network and these are, for example, local clustering coefficient or number of user's acquaintances. The local clustering coefficient describes to what extend the neighbours of a given user create a clique i.e. fully connected graph. Local clustering coefficient was presented by Duncan J. Watts and Steven Strogatz who used it to investigate whether a given graph is a small-world network[4]. User degree centrality in the case of undirected networks, it is expressed by the number of relationships that one has. In directed networks, both the indegree and outdegree centrality can be measured. The former takes into account the number of members that are adjacent to a particular member of the community[2]. The latter takes into account the number of outgoing relations of a given user[31,32]. It should be emphasized that none of the described above methods can be in a straightforward manner used in the multi-layered environment. Thus, there is a need to redefine these concepts for structures on which more than one relation can be defined.

## 3. Multi-layered Social Network

The first step, before the analysis of users' neighbourhoods can be performed, is to define what structure will be used to represent the gathered data. As it was mentioned before, many types of relationships can exist within one social network thus the natural representation of such data is a set of graphs that share common set nodes and relations in a single graph reflect one type of connections. In this paper each of these graphs is called a layer (representing one type of relation) and the whole concept is called a multi-layered social network $MSN$ is defined as a tuple $<V, E, L>$ where: $V$ – is a not-empty set of nodes (human entities); $E$ – is a set of tuples $<x,y,l>$, $x,y \in V$, $l \in L$, $x \neq y$ and for any two tuples $<x,y,l>$, $<x',y',l'> \in E$ if $x=x'$ and $y=y'$ then $l \neq l'$; $L$ – is a set of distinct layers.

Tuple $<x,y,l>$ is an edge $e$ from $x$ to $y$ on the layer $l$ in the multi-layered social network ($MSN$). The assumption is made that the connections from $x$ to $x$ (loops) are not present in the network, i.e. that $x \neq y$. In addition, only one edge $e=<x,y,l>$ from $x$ to $y$ exists on a given layer $l$. The maximum number of edges that can exist between users $x$ and $y$ equals $|L|$. Note that edges in the $MSN$ are directed i.e. $<x,y,l> \neq <y,x,l>$.

Each layer corresponds to one type of relationship between users[8]. Different relationships can result from the character of connections, types of communication channel, or types of activities that users can perform within a given system. The examples of relationships resulting from their different character are e.g.

friendship, family or work ties. Different communication channels that result in different types of connections are email, VoIP, instant messenger, etc. The separate relationship types can be also defined based on users' activities such as publishing photos, commenting photos, adding photos to favourites, and others[8]. The last of the enumerated types of relations possess a semantic meaning as for example publishing photos is a much more proactive action than just adding photos to favourites. Another example where information about users' activities has a semantic meaning can be forum where people who are very active and post a lot of queries can be perceived as new to a field. On the other hand, people who comment a lot but do not post any queries can be seen as experts in a field.

Set of nodes *V* and edges *E* from only one layer $l \in L$ correspond to a simple, one-layered social network $<V, E, \{l\}>$. A multi-layered social network $MSN=<V,E,L>$ may be represented by a directed multi-graph. Hence, all the below proposed structural measures can also be applied to other kinds of complex networks that are described by means of multi-graphs.

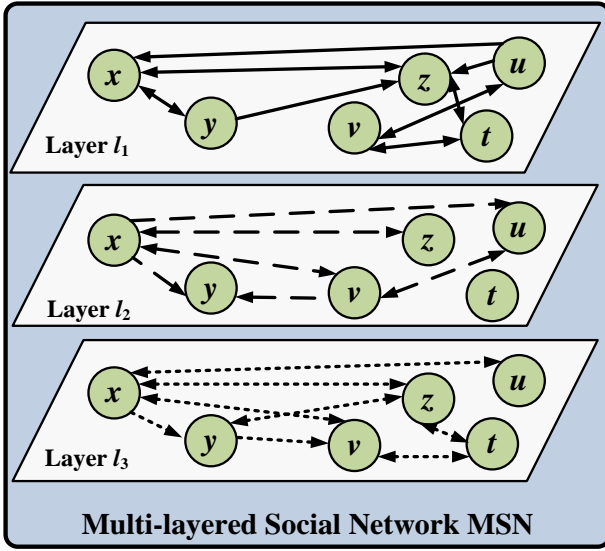

Fig. 1. An example of the multi-layered social network MSN

In order to graphically present the concept of MSN the example of three-layered social network is shown in Figure 1. The set of nodes consists of $\{t, u, v, x, y, z\}$ so there are six users in the network that can be connected with each other's on three layers: $l_1$, $l_2$ and $l_3$. On the layer $l_1$, eight relationships (tuples) between users: $<x,y,l_1>$, $<y,x,l_1>$, $<x,z,l_1>$, $<z,x,l_1>$, $<y,z,l_1>$, $<u,z,l_1>$, $<u,v,l_1>$, $<v,u,l_1>$ can be distinguished. Relationships on the layers $l_2$ and $l_3$ are defined in the same manner.

## 4. Multi-layered Neighbourhood

Neighbourhood $N(x,l)$ of a given node *x* on a given layer *l* for multi-layered social network $MSN=<V,E,L>$ is defined as:

$$N(x,l) = \left\{ y : <y,x,l> \in E \lor <x,y,l> \in E \right\} \quad (1)$$

Set $N(x,l)$ is equivalent to simple neighbourhood for regular one-layered social networks.

Table 1. Node neighbourhoods for each layer for MSN from Figure 1

| Node | Layer l1 | Layer l2 | Layer l3 |
|---|---|---|---|
| *x* | $\{u,y,z\}$ | $\{u,v,y,z\}$ | $\{u,v,y,z\}$ |
| *y* | $\{x,z\}$ | $\{v,x\}$ | $\{v,x,z\}$ |
| *z* | $\{t,u,x,y\}$ | $\{x\}$ | $\{t,x,y\}$ |
| *u* | $\{v,x,z\}$ | $\{v,x\}$ | $\{x\}$ |
| *t* | $\{v,z\}$ | $\{\}$ | $\{v,z\}$ |
| *v* | $\{t,u\}$ | $\{u,x,y\}$ | $\{t,x,y\}$ |

Multi-layered neighbourhood of a given node *x* with a minimum number of layers required − $\alpha$, $1 \leq \alpha \leq |L|$, is a set of nodes, which are neighbours of node *x* on at least $\alpha$ layers in the MSN. Five different versions of multi-layered neighbourhood may be distinguished.

The first one is the multi-layered neighbourhood $MN^{In}(x,\alpha)$ derived from the edges incoming to node *x*, in the following way:

$$MN^{In}(x,\alpha) = \left\{ y : \left| <y,x,l> \in E \right| \geq \alpha \right\} \quad (2)$$

The value of $MN^{In}(x,\alpha)$ denotes the set of neighbours that are connected to node *x* with at least $\alpha$ edges, i.e. on at least $\alpha$ layers of MSN. For $\alpha=1$, we need an edge on only one layer, while for $\alpha=|L|$, if node $y \in MN^{In}(x,\alpha)$, then user *y* must have edges to a given node *x* on all existing layers. For the example MSN from Figure 1, $MN^{In}(x,1)=\{u,v,y,z\}$, $MN^{In}(x,2)=\{u,v,z\}$, $MN^{In}(x,3)=\{z\}$.

Another multi-layered neighbourhood $MN^{Out}(x,\alpha)$ respects only edges outgoing from node *x*.

$$MN^{Out}(x,\alpha) = \left\{ y : \left| <x,y,l> \in E \right| \geq \alpha \right\} \quad (3)$$

For the MSN from Figure 1, we have $MN^{Out}(x,1)=\{u,v,y,z\}$, $MN^{Out}(x,2)=\{u,v,y,z\}$, $MN^{Out}(x,3)=\{y,z\}$.

If we consider incoming or outgoing edges on any layers, then we obtain $MN^{InOutAny}(x,\alpha)$:

$$MN^{InOutAny}(x,\alpha) = \left\{ \begin{array}{l} y : \left| <x,y,l> \in E \right| \geq \alpha \land \\ \left| <y,x,l> \in E \right| \geq \alpha \end{array} \right\} \quad (4)$$

Neighbourhood $MN^{InOutAny}(x,\alpha)$ includes nodes that have at least $\alpha$ incoming and $\alpha$ outgoing edges to and from node *x*, respectively, but these edges may occur on different layers. For the network from Figure 1, there are following sets: $MN^{InOutAny}(x,1)=\{u,v,y,z\}$, $MN^{InOutAny}(x,2)=\{u,v,z\}$ $MN^{InOutAny}(x,3)=\{z\}$.

The next type of multi-layered neighbourhood $MN^{InOut}(x,\alpha)$ is quite similar to $MN^{InOutAny}(x,\alpha)$ but it is more restrictive. Each neighbour $y \in MN^{InOut}(x,\alpha)$ must have bidirectional connections on at least $\alpha$ layers in MSN, i.e. both the incoming and outgoing edge have to occur on the same layer to satisfy the condition, as follows:

$$MN^{InOut}(x,\alpha) = \left\{ y : \left| \left\{ l : \begin{array}{l} < x,y,l >\in E \\ \wedge < y,x,l >\in E \end{array} \right\} \right| \geq \alpha \right\} \quad (5)$$

In the example MSN, Figure 1, we have $MN^{InOut}(x,1)=\{u,v,y,z\}$, $MN^{InOut}(x,2)=\{v,z\}$ $MN^{InOut}(x,3)=\{z\}$.

The final, fifth neighbourhood $MN(x,\alpha)$ is the least restrictive. It takes into consideration, any incoming or outgoing edges on any layer but the total number of these layers should be at least $\alpha$:

$$MN(x,\alpha) = \left\{ y : \left| \left\{ \begin{array}{l} l :< x,y,l >\in E \\ \vee < y,x,l >\in E \end{array} \right\} \right| \geq \alpha \right\} \quad (6)$$

In the example social network, the neighbourhoods are as follows: $MN(x,1)=\{u,v,y,z\}$, $MN(x,2)=\{u,v,y,z\}$ $MN(x,3)=\{u,y,z\}$, see Figure 1. $MN(x,\alpha)$ is utilized for studies described in further sections of this paper. However, the other neighbourhoods may also be used and it would require only small modifications of the proposed below measures.

Table 2. Multi-layered neighbourhoods for the nodes from Figure 1

| Node | MN(•,1) | MN(•,2) | MN(•,3) |
|------|---------|---------|---------|
| *x* | {u,v,y,z} | {u,v,y,z} | {u,y,z} |
| *u* | {x,v,z} | {x,v} | {x} |
| *z* | {t,u,x,y} | {t,x,y} | {x} |
| *u* | {z,v,x} | {x,v} | {x} |
| *t* | {v,z} | {v,z} | {} |
| *v* | {t,u,x,y} | {t,u,x,y} | {} |

Note that $MN^{InOutAny}(x,\alpha) = MN^{In}(x,\alpha) \cap MN^{Out}(x,\alpha)$ and $MN^{InOut}(x,\alpha) \subseteq MN^{InOutAny}(x,\alpha) \subseteq MN(x,\alpha)$, $MN^{InOut}(x,\alpha) \subseteq MN^{In}(x,\alpha) \subseteq MN(x,\alpha)$ and $MN^{InOut}(x,\alpha) \subseteq MN^{Out}(x,\alpha) \subseteq MN(x,\alpha)$. The smallest neighbourhood is $MN^{InOut}(x,\alpha)$ while the largest is $MN(x,\alpha)$. It means that $MN^{InOut}(x,\alpha)$ is the most restrictive whereas $MN(x,\alpha)$ is the least.

Although multi-layered neighbourhood is a structural measure, as it was presented in Sec. 2 and 3, it also has semantic meaning. For example, a large $MN(x,\alpha)$ for a large $\alpha$ will be an indicator that person $x$ is a communication hub. $MN^{In}(x,\alpha)$ is more restrictive, so it provides some more detailed semantic meaning. Its large value when $\alpha$ is large means that there is a lot of incoming interaction towards $x$. For example, in the context of the company, it may mean that $x$ is a line manager or another person to whom a group of people reports using different communication channels. These channels can be treated as separate layers in the network. On the other hand, a high value of $MN^{Out}(x,\alpha)$ for large $\alpha$ means that person $x$ is responsible for propagating information in the network, e.g. a person responsible for bulletin in the organisation. It also helps to investigate which communication channels are neglected. Such examples can be multiplied. Please note that it can serve to investigate both positive and negative human behaviours. For example, a person within the company with high $MN^{Out}(x,\alpha)$ who is not responsible for propagating information can be seen as a "chatterbox" who spends more time on communication than on doing the job.

Also at the level of the whole network, the multi-layered neighbourhood can be analysed. The power-law distribution of $MN(x,\alpha)$ depending on $\alpha$ means that not all types of relations are fully used and people tend to focus on only few relation types and neglect the others.

## 5. Cross-layer Clustering Coefficient

A cross-layer clustering coefficient $CLCC(x,\alpha)$ was introduced to allow calculation of the clustering coefficient for the multi-layered social network MSN. For a given node $x$, and $x$'s non-empty neighbourhood $MN(x,\alpha)$, cross-layer clustering coefficient $CLCC(x,\alpha)$ is computed in the following way:

$$CLCC(x,\alpha) = \frac{\displaystyle\sum_{l \in L} \sum_{y \in MN(x,\alpha)} \left( \begin{array}{l} in\left(y,MN(x,\alpha),l\right) \\ + out\left(y,MN(x,\alpha),l\right) \end{array} \right)}{2 \cdot \left| MN(x,\alpha) \right| \cdot \left| L \right|} \quad (7)$$

where: $in(y,MN(x,\alpha),l)$ – the weighted indegree of node $y$ in the multi-layered neighbourhood $MN(x,\alpha)$ of node $x$ within the simple one-layered network $<V, E, \{l\}>$, i.e. within only one layer $l$; $out(y,MN(x,\alpha),l)$ – the weighted outdegree of node $y$ in the multi-layered neighbourhood $MN(x,\alpha)$ of node $x$ in the network containing only one layer $l$.

If neighbourhood $MN(x,\alpha)=\varnothing$, then $CLCC(x,\alpha)=0$.

The weighted indegree $in(y, MN(x,\alpha),l)$ for a given node $y$ in the network $<V, E, \{l\}>$ containing one layer $l$ is the sum of all weights $w(z,y,l)$ of edges $<z,y,l>$ incoming to node $y$ from other nodes $z$ that are from layer $l$ and belong to multi-layered neighbourhood $MN(x,\alpha)$:

$$in\left(y,MN(x,\alpha),l\right) = \sum_{z \in MN(x,\alpha)} w(z,y,l). \quad (8)$$

Likewise, the weighted outdegree $out(y,MN(x,\alpha),l)$ for a given node $y$ and neighbourhood $MN(x,\alpha)$ is the sum of all weights $w(y,z,l)$ of the outgoing edges $\langle y,z,l \rangle$ that come from $y$ to $x$'s neighbours $z$ on layer $l$:

$$out\left(y, MN\left(x,\alpha\right),l\right) = \sum_{z \in MN(x,\alpha)} w\left(y,z,l\right). \qquad (9)$$

Note that if the sum of weights of outgoing edges for a given node x is 1 (this is the usual practice in social network analysis[33], then $CLCC(x,\alpha)$ is always from the range [0;1]. The value $CLCC(x,\alpha)=1$, if each neighbour $y \in MN(x,\alpha)$ has outgoing relationships towards all other nodes $z \in MN(x,\alpha)$ and only to them. The value of $CLCC(x,\alpha)$ equals 0 when $x$ has only one neighbour, i.e. $|MN(x,\alpha)|=1$, or when none of the $x$'s neighbours $y \in MN(x,\alpha)$ has any relationship with any neighbour $z \in MN(x,\alpha)$.

For node $t$ from the example social network, Figure 1, $MN(t,1)=MN(t,2)=\{v,z\}$ but $CLCC(t,1)=CLCC(t,2)=0$ because there are no edges between $v$ and $z$. Due to $MN(t,3)=\{\}$ we have $CLCC(t,3)=0$. For node $z$: $MN(z,1)=\{t,u,x,y\}$, $MN(z,2)=\{t,x,y\}$, $MN(z,3)=\{x\}$. If weights of all edges equal 1, then $CLCC(z,1)={}^{16}/_{24}={}^{2}/_{3}$ and $CLCC(z,2)={}^{8}/_{18}={}^{4}/_{9}$. Since there is only one neighbour $x$ in $MN(z,3)$, the value $CLCC(z,3)=0$.

The formula for another measure – multi-layered clustering coefficient ($MCC$) as well as the two special cases of cross-layer clustering coefficient $CLCC(x,\alpha)$ were described in Ref. 34. One of them is multi-layered clustering coefficient in extended neighbourhood ($MCCEN$). It is, in fact, the cross-layer clustering coefficient for only one layer($\alpha=1$), i.e. $MCCEN(x)=CLCC(x,1)$. Multi-layered clustering coefficient in reduced neighbourhood ($MCCRN$) presented in Ref. 34 is equivalent to the cross-layer clustering coefficient for all layers, $\alpha=|L|$ i.e. $MCCRN(x)=CLCC(x,|L|)$.

Cross-Layered Clustering Coefficient can also be interpreted in the semantic context. For example, people who are in the professional relationships (e.g. co-workers) prefer to communicate via e-mail as it enables to keep track of their interactions. It means that their clustering coefficient will be higher at one layer (assuming that their neighbours communicate with each other) than at the rest of the layers. A different situation occurs in more private social circumstances where people tend to use different communication channels and it can result in higher value of $CLCC$, although the clustering coefficient for a single layer may remain not particularly high. It is a consequence of the fact that our friends (neighbours in the network) can interact with each other using different communication layers (phone call, e-mail, text messages, etc.).

## 6. Multi-layered Degree Centrality

Apart from clustering coefficient, there exists another measure commonly used in social network analysis – degree centrality, Sec. 6.1. Some of its versions for multi-layered social network, presented in Sec. 6.2, were initially studied in Ref. 35. However, in this paper, we also introduce some other concepts for multi-layered degree centralities, see Sec. 6.3-6.5.

### 6.1. Regular Degree Centrality

For the regular, one-layered social network, degree centrality $DC(x)$ for node $x$ is defined as follows:

$$DC(x) = \frac{d(x)}{m-1} \qquad (10)$$

where: $d(x)$ – the number of the first level neighbours that are connected with $x$ either with incoming or outgoing edge; $m$ – the total number of members in the social network, $m=|V|$.

Indegree centrality $IDC(x)$ for node $x$, in turn, takes into account only edges incoming to node $x$, in the following way:

$$IDC(x) = \frac{i(x)}{m-1} \qquad (11)$$

where: $i(x)$ – the number of the first level neighbours that are connected to $x$ with edges directed from these neighbours to $x$.

Another measure is outdegree centrality $ODC(x)$ that respects only edges outgoing from node $x$:

$$ODC(x) = \frac{o(x)}{m-1} \qquad (12)$$

where: $o(x)$ – the number of the first level neighbours $y$ of node $x$, for which exist edges from $x$ to $y$.

Note that in the case of weighted one-layer social network it is possible to use sum of edges weights between $x$ and its neighbours instead of number of the first level neighbours.

### 6.2. Cross-layer Degree Centrality

The first multi-layered degree centrality is called cross-layer degree centrality ($CDC$). It is defined as a sum of edge weights both incoming to and outgoing from node $x$ towards multi-layered neighbourhood $MN(x,\alpha)$ divided by the number of layers and total network members:

$$CDC(x,\alpha) = \frac{\sum\limits_{y \in MN(x,\alpha)} w(x,y,l) + \sum\limits_{y \in MN(x,\alpha)} w(y,x,l)}{(m-1)\,|\,L\,|} \quad (13)$$

where: $w(x,y,l)$ – the weight of edge $\langle x,y,l \rangle$.

Similarly to different versions of degree centrality $DC(x)$ - $IDC(x)$ and $ODC(x)$ (see Sec. 4.1), we can define cross-layer indegree centrality $CDC^{In}(x,\alpha)$ in the multi-layered social network MSN:

$$CDC^{In}(x,\alpha) = \frac{\sum\limits_{y \in MN(x,\alpha)} w(y,x,l)}{(m-1)\,|\,L\,|} \quad (14)$$

and cross-layer outdegree centrality $CDC^{Out}(x,\alpha)$:

$$CDC^{Out}(x,\alpha) = \frac{\sum\limits_{y \in MN(x,\alpha)} w(x,y,l)}{(m-1)\,|\,L\,|} \quad (15)$$

As in the case of the cross-layer clustering coefficient $CLCC(x,\alpha)$, the value of $CDC(x,\alpha)$ directly depends on the parameter $\alpha$, which determines the multi-layered neighbourhood of a given social network member $x$.

### 6.3. Multi-layered Degree Centrality Version 1

The other three multi-layered degree centralities are not calculated based on $MN(x,\alpha)$ but using the local neighbourhood in particular layer $N(x,l)$, (see Sec. 4). The first of them $MDC^{(1)}(x)$ is defined as a sum of $x's$ local weighted degree centralities in each layer $l$ divided by the number of layers:

$$MDC^{(1)}(x) = \frac{\sum\limits_{l \in L}\left(\sum\limits_{y \in N(x,l)} w(x,y,l) + \sum\limits_{y \in N(x,l)} w(y,x,l)\right)}{(m-1)\,|\,L\,|} \quad (16)$$

The first multi-layered indegree centrality $MDC^{(1)In}(x)$ is defined as follows:

$$MDC^{(1)In}(x) = \frac{\sum\limits_{l \in L}\sum\limits_{y \in N(x,l)} w(y,x,l)}{(m-1)\,|\,L\,|} \quad (17)$$

and the first multi-layered outdegree centrality $MDC^{(1)Out}(x)$:

$$MDC^{(1)Out}(x) = \frac{\sum\limits_{l \in L}\sum\limits_{y \in N(x,l)} w(x,y,l)}{(m-1)\,|\,L\,|} \quad (18)$$

### 6.4. Multi-layered Degree Centrality Version 2

The next multi-layered degree centrality $MDC^{(2)}(x)$ is a sum of $x's$ local weighted degree centralities in each layer but in opposite to $MDC^{(1)}(x)$, it is divided by the quantity of the union of $x's$ neighbourhoods from all layers.

$$MDC^{(2)}(x) = \frac{\sum\limits_{l \in L}\left(\sum\limits_{y \in N(x,l)} w(x,y,l) + \sum\limits_{y \in N(x,l)} w(y,x,l)\right)}{(m-1)\,|\,MN(x,1)\,|} \quad (19)$$

Note that the union of neighbourhood sets from all layers for a given member $x$ is the same as the multi-layered neighbourhood $MN(x,\alpha)$ for $\alpha=1$, i.e. $\bigcup_{l \in L} N(x,l) = MN(x,1)$

The multi-layered indegree centrality $MDC^{(2)In}(x)$ is:

$$MDC^{(2)In}(x) = \frac{\sum\limits_{l \in L}\sum\limits_{y \in N(x,l)} w(y,x,l)}{(m-1)\,|\,MN(x,1)\,|} \quad (20)$$

and the first multi-layered outdegree centrality $MDC^{(2)Out}(x)$:

$$MDC^{(2)Out}(x) = \frac{\sum\limits_{l \in L}\sum\limits_{y \in N(x,l)} w(x,y,l)}{(m-1)\,|\,MN(x,1)\,|} \quad (21)$$

### 6.5. Multi-layered Degree Centrality Version 3

The last multi-layered degree centrality $MDC^{(3)}(x)$ is quite similar to $MDC^{(2)}(x)$ but instead of the neighbourhood sets from all layers, the sum of $x's$ local weighted degree centralities in each layer is divided by the sum of neighbourhood quantities on each layer:

$$MDC^{(3)}(x) = \frac{\sum\limits_{l \in L}\left(\sum\limits_{y \in N(x,l)} w(x,y,l) + \sum\limits_{y \in N(x,l)} w(y,x,l)\right)}{(m-1)\sum\limits_{l \in L}|\,N(x,l)\,|} \quad (22)$$

The third version of first multi-layered indegree centrality $MDC^{(3)In}(x)$:

$$MDC^{(3)In}(x) = \frac{\sum_{l \in L} \sum_{y \in N(x,l)} w(y,x,l)}{(m-1)\sum_{l \in L}|N(x,l)|} \quad (23)$$

and the third version of first multi-layered outdegree centrality $MDC^{(3)Out}(x)$:

$$MDC^{(3)Out}(x) = \frac{\sum_{l \in L} \sum_{y \in N(x,l)} w(x,y,l)}{(m-1)\sum_{l \in L}|N(x,l)|} \quad (24)$$

Both Cross-Layered and Multi-Layered Degree Centralities are helpful in the interpretation of the social network semantics. They provide some more information than $MN(x,\alpha)$ as they take into consideration not only the number of relations but also their quality, i.e. connection strength.

## 7. Experimental Study and Discussion

The real-world dataset used for experiments contains information about activities performed by users on the forum existing within the web site *extradom.pl* (equivalent to 'extraordinary house'). The portal gathers people, who are engaged in building their own houses in Poland. It helps them to exchange best practices, experiences, evaluate various constructing projects and technologies or simply to find the answers to their questions provided by others. The data comes from the period of 7 months (for the experiments in Sec. 7.3 an extended period was analysed). The Multi-Layered Social Network (*MSN*) was created using the multi-layered social network creation process described in Ref. 33. It consisted of 4,404 users and 11 different layers. All layers and the number of relationships in each of them are presented in Table 3. The different layers were extracted based on various user activities within the forum: creation of the forum itself, management of topic groups, creation of new topics, subscription of topics, posting and post commenting. These activities link their owners, i.e. if user $x$ comments the post of user $y$ they both get into common relationship (layer 11). In another example, two users $x$ and $y$ can add their own topics to the same topic group (layer 2), they can also subscribe the same topic (layer 6) or comment the same post (layer 9), hence, they 'meet' at that topic group, topic or post, respectively. These 'meetings' can be more or less frequent, hence, we may have stronger or weaker mutual relationship.

Overall, the more common activities (creation of topics, posts, comments, subscriptions) the stronger the relationship. The details and formulas referring MSN creation can be found in Ref. 33 and Ref. 8.

The greatest number of relations is for layers number 6, 8 and 9 and it equals 286,502 whereas on layer number 10 there are only 39 relations. This shows the great diversity in user activities and their frequencies in the investigated system.

Layers themselves provide some semantic knowledge and the number of relations within these layers indicates community global interests towards specific activities and the amount of engagement that people are keen to devote to a given social network as different layers require different level of involvement. For example, within the investigated network, posting and further commenting these posts by community is not very common activity.

Table 3. The number of relationships in layers within the MSN

| Layer index | Relationships on the layer derived from two kinds of user activities linking users in the forum | No. of relations |
|---|---|---|
| 1 | Forum creation activity - TopicGroup group addition | 265 |
| 2 | Topic addition - Topic addition | 33,308 |
| 3 | Forum creation activity - Topic subscribing | 4,358 |
| 4 | TopicGroup group addition - Topic subscribing | 5,034 |
| 5 | Topic addition - Topic subscribing | 92,959 |
| 6 | Topic subscribing - Topic subscribing | 286,502 |
| 7 | Topic addition - Post posting | 92,959 |
| 8 | Topic subscribing - Post posting | 286,502 |
| 9 | Post posting - Post posting | 286,502 |
| 10 | Forum creation activity - Post commenting | 39 |
| 11 | Post posting - Post commenting | 2,334 |

### 7.1. Static Analysis

In the first stage of the experiments, the multi-layered neighbourhood $MN(x,\alpha)$ was calculated for each user $x$. The calculations were performed separately for 11 different values of parameter $\alpha$, i.e. 1, ..., 11.

In Figure 2 the numbers of unique users within multi-layered neighbourhoods $|MN(x,\alpha)|>0$ for different values of $\alpha$ are presented with the black columns. It can be noticed that for the values of $\alpha$ from the set $\{1,...,6\}$, the number of users with non-empty multi-layered neighbourhoods is high, over 96% of all users. For $\alpha =1$ – none of the users $x$ has $MN(x,\alpha)=\{\}$ and even for $\alpha=6$ only 164 users have empty multi-layered neighbourhoods. The rapid drop for $\alpha=7$ is caused by the fact that there are 6 layers in which there is a large number of relationships (layers no. 2, 5, 6, 7, 8 and 9) and the rest of the layers are very sparse, with the considerably smaller number of connections (layers 1, 3, 4 and 10).

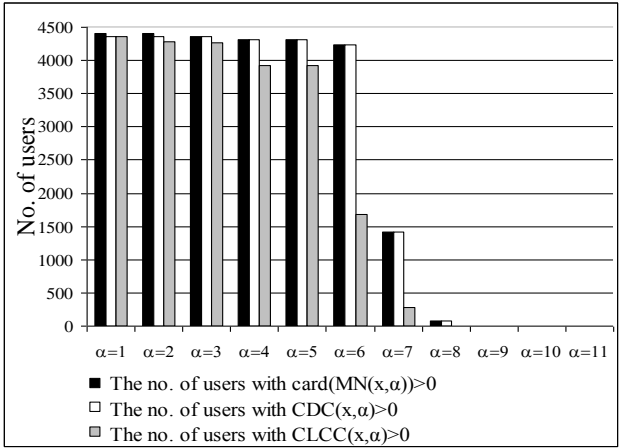

Fig. 2. The number of users in multi-layered neighbourhoods $MN(x,\alpha)$ for different $\alpha$ values

Next, the cross-layer clustering coefficient and cross-layer degree centrality were calculated. Both measures were evaluated separately for 11 different values of parameter $\alpha$ *i.e.* 1, ..., 11. Similarly to $MN(x,\alpha)$, the number of non-zero values of $CDC(x,\alpha)$ and $CLCC(x,\alpha)$ for different $\alpha$ are presented (white and grey columns respectively in Figure 2).

It can be noticed that for the min. number of layers from 1 to 6 required, the number of users with non-zero $CDC(x,\alpha)$ is high, over 96% of all users. For $\alpha=1$, $CDC$ of only 40 users equal zero and even for $\alpha=6$ only 164 decrease can be noticed for $\alpha=6$ not for $\alpha=7$ as in $MN(x,\alpha)$ and $CDC(x,\alpha)$. For $\alpha=5$, there are 89% of users and for $\alpha=6$ only 38.1% users with $CLCC(x,\alpha)>0$ and for each $\alpha$ there are more users with $CLCC(x,\alpha)=0$ than with $CDC(x,\alpha)=0$ what shows that there exist people who have acquaintances but these friends are not in relationships with each other at all. The phenomena that we observe for $\alpha=6$ is not typical in social networks where people tend to create clusters ("friend of my friend is my friend"). If a user has a small clustering coefficient but a lot of relations, then the semantic meaning of relation can be assigned to this, i.e. that his relationships are rather for finding information rather than for developing social life.

Table 4. Parameters' values for correlation function used for fitting distribution of the multi-layered degree centrality

| Measure | $A$ | $T$ | $CR$ |
|---------|-----|-----|------|
| $MDC^{(1)}$ | 0.0004 | -0.0004 | 0.348 |
| $MDC^{(2)}$ | 0.0004 | -0.001 | 0.942 |
| $MDC^{(3)}$ | 6E-05 | -0.001 | 0.965 |

Finally, all three versions of multi-layered degree centrality were calculated. First, the distributions of all multi-layered degree centralities were analysed. As a result of the fitting process, we obtained the function that approximates the experimental data in the best way. This is an exponential decay function described by the following formula, see also Figure 3:

$$MDC^{(i)}(x) = A \cdot e^{x/t}. \qquad (25)$$

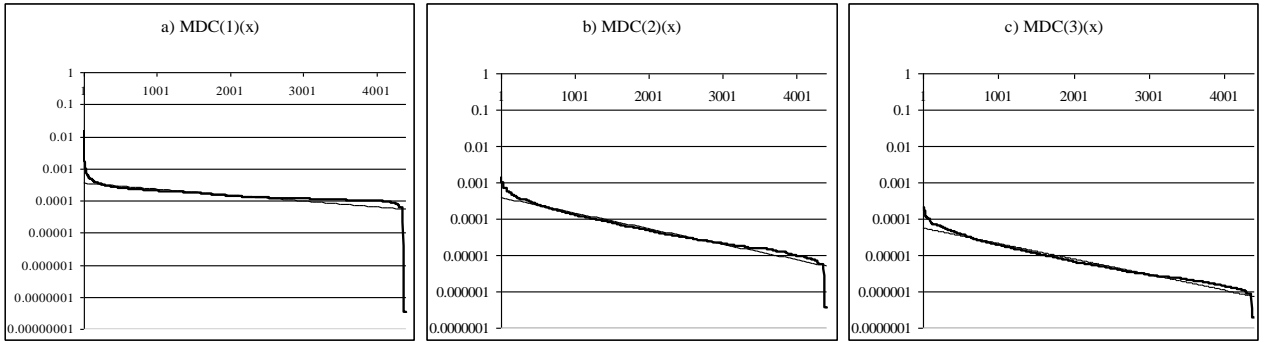

Fig. 3. Values of MDCs (a thick line) and the fitting functions (a thin line) for particular users (X axis)

users have $CDC=0$. The rapid drop for $\alpha=7$ is caused by the same reason as in the case of $MN(x,\alpha)$. For $CLCC(x,\alpha)$, the tendency is similar. However, the rapid

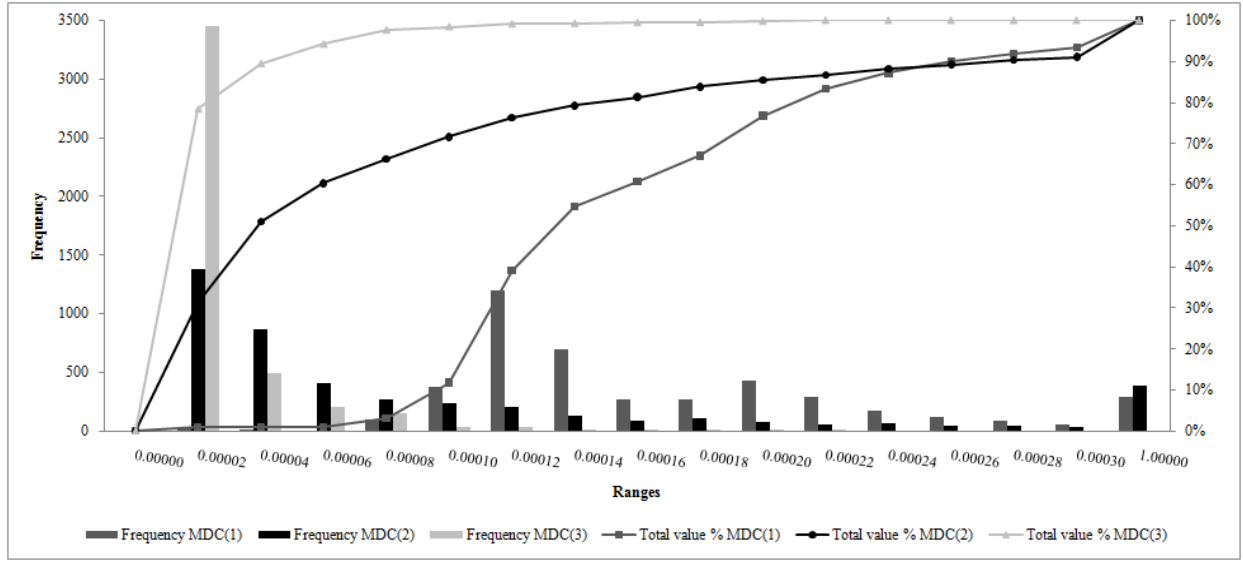

Fig. 4. Histogram and accumulative percentage distribution of $MDC^{(1)}$, $MDC^{(2)}$, $MDC^{(3)}$

Table 5 Histogram and accumulative percentage distribution of $MDC^{(1)}$, $MDC^{(2)}$, $MDC^{(3)}$ for given ranges

| Ranges | Frequency $MDC^{(1)}$ | Total value % $MDC^{(1)}$ | Frequency $MDC^{(2)}$ | Total value % $MDC^{(2)}$ | Frequency $MDC^{(3)}$ | Total value % $MDC^{(3)}$ |
|---|---|---|---|---|---|---|
| 0.00000 | 0 | 0.00% | 0 | 0.00% | 0 | 0.00% |
| 0.00002 | 40 | 0.91% | 1379 | 31.31% | 3451 | 78.36% |
| 0.00004 | 1 | 0.93% | 870 | 51.07% | 489 | 89.46% |
| 0.00006 | 0 | 0.93% | 408 | 60.33% | 209 | 94.21% |
| 0.00008 | 101 | 3.22% | 266 | 66.37% | 151 | 97.64% |
| 0.00010 | 378 | 11.81% | 233 | 71.66% | 36 | 98.46% |
| 0.00012 | 1200 | 39.06% | 204 | 76.29% | 30 | 99.14% |
| 0.00014 | 692 | 54.77% | 131 | 79.27% | 6 | 99.27% |
| 0.00016 | 266 | 60.81% | 93 | 81.38% | 8 | 99.46% |
| 0.00018 | 275 | 67.05% | 108 | 83.83% | 2 | 99.50% |
| 0.00020 | 428 | 76.77% | 75 | 85.54% | 11 | 99.75% |
| 0.00022 | 288 | 83.31% | 55 | 86.78% | 11 | 100.00% |
| 0.00024 | 172 | 87.22% | 68 | 88.33% | 0 | 100.00% |
| 0.00026 | 122 | 89.99% | 43 | 89.31% | 0 | 100.00% |
| 0.00028 | 89 | 92.01% | 44 | 90.30% | 0 | 100.00% |
| 0.00030 | 59 | 93.35% | 35 | 91.10% | 0 | 100.00% |
| 1.00000 | 293 | 100.00% | 392 | 100.00% | 0 | 100.00% |

For each *MDC* version, the values of *A*, *t*, $n_0$ and correlation rate are in Table 4. The correlation rate is high for $MDC^{(2)}$ and $MDC^{(3)}$ what means that exponential decay function approximates the distribution of *MDC* values with a very high accuracy; the highest possible is 1. Low CR for $MDC^{(1)}$ is mainly caused by big differences for the smallest and the highest values of $MDC^{(1)}$.

In Figure 4 and Table 5, the histogram and percentage distribution of all three MDCs values was presented which confirm earlier observations on MDCs distribution.

## 7.2. *Dynamic Analysis*

The next part of the experiments was focused on the dynamics of the multi-layered neighbourhood $MN(x,\alpha)$. The aim of this part was to investigate how active were users in specific periods of time and on which layers. The study was performed once again on data from *extradom.pl*. This time it was a bigger data set from 17 months period (a shorter 7-month data was used in Sec. 7.1 and 7.2). Within this period there were 23,429 active system users were identified, i.e. those who performed at least one activity at the forum existing within the portal.

First, the data was split into 5 non-overlapping time windows (W1, W2, W3, W4, W5), each covered 90 days. Five different MSNs were created using the multi-layered social network creation process described in Ref. 33 and briefly explained in Sec. 7.1, one MSN for each time window.

Due to the limitations in computational power only five layers were taken into consideration while calculating neighbourhoods in each window. These layers are: Forum:Topic subscribing - Forum:Topic subscribing, Forum:Topic subscribing - Forum:Post posting, Forum:Post posting - Forum:Post posting, Forum:Topic subscribing - Forum:Post commenting, Forum:Post posting - Forum:Post commenting.
For each window the calculations were made separately for $\alpha=1,\ldots,5$.

Table 6 Number of active people for each $\alpha$ value and for different windows and windows combination

| Layer name | $\alpha=1$ | $\alpha=2$ | $\alpha=3$ | $\alpha=4$ | $\alpha=5$ |
|---|---|---|---|---|---|
| Number of people who were not active at all | | | | | |
| Number of no active users | 0 | 5019 | 5019 | 15016 | 19249 |
| Number of people active in all windows for a given $\alpha$ | | | | | |
| W12345 | 306 | 194 | 194 | 6 | 1 |
| Number of people active in four windows for a given $\alpha$ | | | | | |
| W1234 | 26 | 18 | 18 | 0 | 0 |
| W1235 | 48 | 30 | 30 | 0 | 0 |
| W1245 | 98 | 70 | 70 | 0 | 0 |
| W1345 | 65 | 60 | 60 | 0 | 0 |
| W2345 | 448 | 277 | 277 | 1 | 1 |
| Number of people active in three windows for a given $\alpha$ | | | | | |
| W123 | 120 | 84 | 84 | 12 | 3 |
| W124 | 29 | 21 | 21 | 4 | 1 |
| W125 | 71 | 56 | 56 | 4 | 1 |
| W134 | 18 | 13 | 13 | 0 | 0 |
| W135 | 0 | 11 | 11 | 0 | 0 |
| W145 | 58 | 64 | 64 | 4 | 1 |
| W234 | 158 | 109 | 109 | 12 | 3 |
| W235 | 205 | 147 | 147 | 16 | 4 |
| W245 | 278 | 217 | 217 | 6 | 1 |
| W345 | 722 | 451 | 451 | 25 | 7 |
| Number of people active in two windows for a given $\alpha$ | | | | | |
| W12 | 459 | 351 | 351 | 99 | 33 |
| W13 | 106 | 95 | 95 | 36 | 12 |
| W14 | 69 | 61 | 61 | 24 | 8 |
| W15 | 55 | 73 | 73 | 23 | 7 |
| W23 | 680 | 506 | 506 | 114 | 40 |
| W24 | 180 | 153 | 153 | 51 | 17 |
| W25 | 203 | 209 | 209 | 36 | 12 |
| W34 | 726 | 533 | 533 | 120 | 42 |
| W35 | 549 | 455 | 455 | 104 | 36 |
| W45 | 1374 | 893 | 893 | 140 | 48 |
| Number of people active in one window for a given $\alpha$ | | | | | |
| W1 | 1420 | 1227 | 1227 | 750 | 376 |
| W2 | 3009 | 2555 | 2555 | 1537 | 788 |
| W3 | 3999 | 3308 | 3308 | 1906 | 979 |
| W4 | 3242 | 2660 | 2660 | 1546 | 807 |
| W5 | 4708 | 3509 | 3509 | 1837 | 952 |

Figure 5 shows that the largest number of active users is when the parameter $\alpha$ equals 1. Most users are active only within a single window. If users are active in two or more time windows then in most cases these windows follow one another. From the semantic analysis point of view, it can be an indicator that the analysed portal is a kind of problem solving engine rather than a social networking site. People tend to ask for help and once they get it, they stop posting on the forum. Because *extradom.pl* is a system for peoples who build their houses or decorating their apartments it seems to be a valid conclusion that people use the forum only when they need help and stop using it when they receive the answers for their questions or finish their houses.

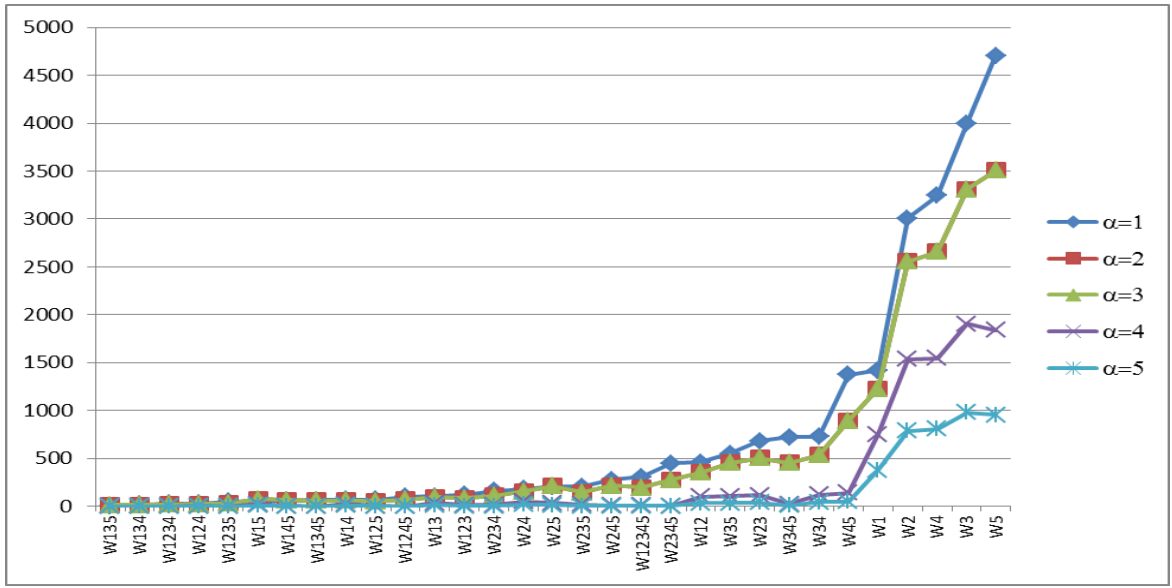

Figure 5 Number of users active in different windows depending on $\alpha$ value

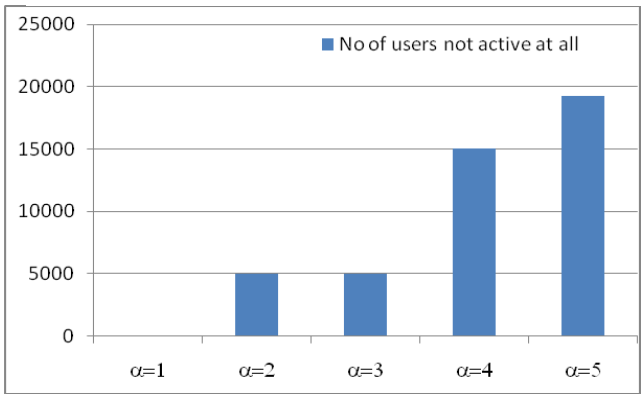

Figure 6 Number of users who are not active at all in relation to $\alpha$ value

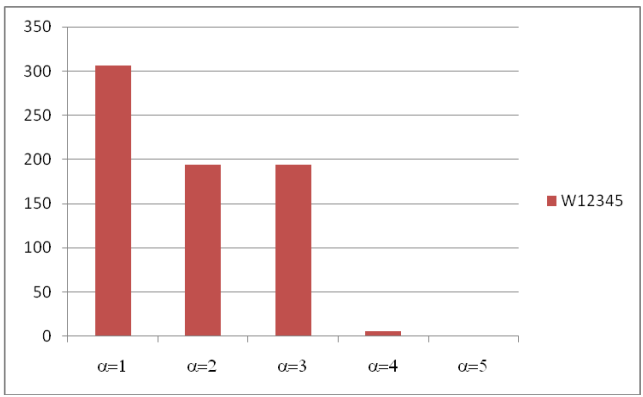

Figure 7 Number of users who are active during the whole analysed time period (W12345) depending on $\alpha$ value

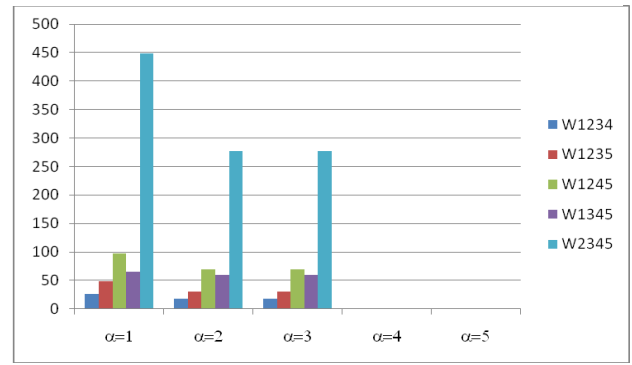

Figure 8 Number of users who are active within four windows out of five depending on $\alpha$ value

In Figure 8 it can be noticed that, when four time windows are taken into consideration, then the largest number of people active within four out of five windows is when W2, W3, W4, and W5 are analysed together. As the data from the investigated system came from the period when the system came into existence (from day one), this can mean that as the system has evolved more and more people have started using it.

Figure 8, Figure 9 and Figure 10 confirm the conclusion that if people are active within more than one time window then these windows follow one after another. Moreover, there are more active people in the later windows and their combinations (W3, W4, W5) than during the first 6 months (W1, W2) of availability of the system.

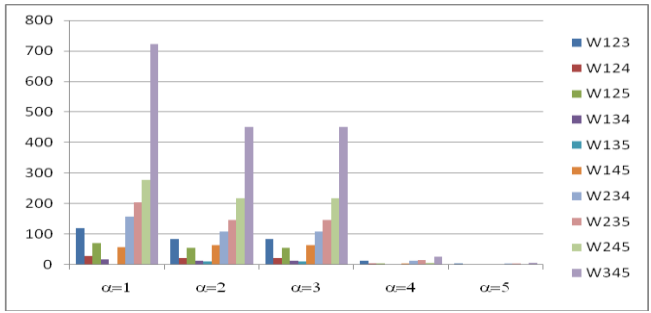

Figure 9 Number of users who are active within three windows out of five in relation to various $\alpha$ values

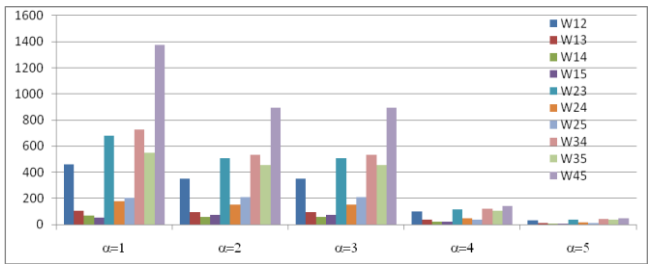

Figure 10 Number of users who are active within two windows out of five depending on $\alpha$ value

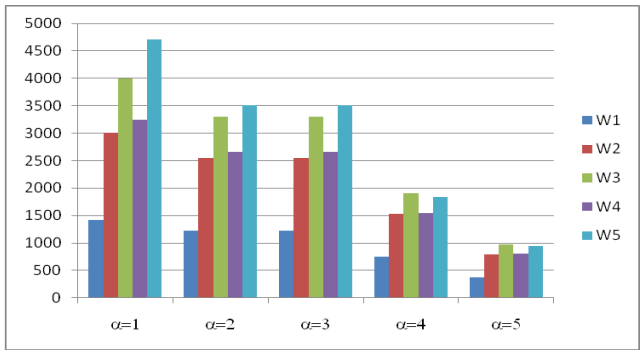

Figure 11 Number of users who are active within one window out of five depending on $\alpha$ value

Another interesting pattern can be noticed when the activity is analysed based on one time window (Figure 11). In all cases in W4 window the number of active people is smaller than in W3 and W5. This shows the seasonality in the works connected with repairs and renovations of the houses as W4 covers months: October, November and December.

## 8. Conclusions

The paper addresses the problem of neighbourhood analysis in multi-layered social networks, in which members can be connected with each other on many different layers. The definition of multi-layered social network and five definitions of multi-layered neighbourhoods (*MN*) have been proposed. The neighbourhoods consist of users whose activities result in relations on at least a given number of layers. Based on *MN* definition, the cross-layer clustering coefficient *CLCC* and cross-layer degree centrality *CDC* have been defined. The former measures the density of connections between neighbours of a given user, while *CDC* determines how strong they are connected to this use. Additionally, three different multi-layered degree centralities *MDC* were proposed. *MDCs* just like *CDC* describe how strong a given user $x$ is connected with $x$'s neighbours but instead of using multi-layered neighbourhoods *MDCs*, it utilizes local neighbourhood on each layer.

The experiments on real web portal revealed that the average size of *MN* generally decreases with the increasing number of layers and this triggers the same behaviour for *CLCC* and *CDC*. It shows that although people are exposed to many different types of relationships, they tend to narrow their activities into just one or two types. Analysis of the multi-layered neighbourhood dynamics disclosed that users tend to be active in only one or two consecutive time windows. Moreover, there are very few people (1.3%), who are active within the whole analysed time period.

Furthermore, we have shown that it is possible to assign semantic meaning to the structural analysis of the network. Both static and dynamic results of the analysis can be interpreted in terms of semantic information that they carry. Of course, this interpretation heavily depends on the context of the analysis so the more additional information one possesses about the network and its components, the better. Such additional information can be, e.g. profiles of the network members or the usage data gathered e.g. using surveys[36]. This study has also revealed that in order to discover semantics of the network and interactions between people, one does not necessary has to perform complicated and resource consuming natural language processing tasks.

The MSN allows analysing not only strength of the relations between people but also their nature. This concept can be applied in telecommunication companies whose customers' behaviour is investigated by analysing different ways of communication, e.g. direct calls, text messages, video conferences, etc. Nowadays, the Customer Relationship Management is highly influenced by social networks[37], so this domain can also benefit from using SNA. The MSN can also support other cooperative actions of users like collaborative Information Retrieval or metadata management, trust management between its members, targeted marketing, recommender systems or collaboration in e-learning systems[38]. An interesting and powerful method that can be utilized in these domains is collective classification, in which labels (classes) are assigned to the network nodes using as an input complex structural network measures like ones proposed in Ref. 39.

**Acknowledgements**

The work was partially supported by The Polish National Science Centre, the research project 2010-13. The publication has been prepared as part of the project of the City of Wrocław, entitled "Green Transfer" - academia-to-business knowledge transfer project co-financed by the European Union under the European Social Fund, under the Operational Programme Human Capital (OP HC): sub-measure 8.2.1.

**References**

[1]     R. Hanneman, and M. Riddle, *Introduction to social network methods, online textbook*, Riverside, CA: University of California, 2005, Available: http://faculty.ucr.edu/~hanneman/nettext/.

[2]     S. Wasserman, and K. Faust, *Social network analysis: Methods and applications*. Cambridge University Press, New York, 1994.

[3]     J. Scott, *Social network analysis: developments, advances, and prospects* Social Network Analysis and Mining  Vol. 1, no 1, pp. 21-26, DOI: 10.1007/s13278-010-0012-6

[4]     D.J. Watts, and S. Strogatz, Collective dynamics of 'small-world' networks, *Nature*, vol. 393, 1998, pp. 440-444.

[5]     B. Entwisle, K. Faust, R.R. Rindfuss, T. Kaneda, Networks and contexts: Variation in the structure of social ties. The American Journal of Sociology, vol. 112, 2007, pp. 1495–1533.

[6]     M. McPherson, L. Smith-Lovin, J.M. Cook, Birds of a feather: Homophily in social networks, Annu Rev Sociol, vol. 27, 2001, pp.415–444.

[7]     J.F. Padgett, C.K. Ansell, Robust action and the rise of the Medici, 14001434, The American Journal of Sociology, vol. 98, 1993, pp. 1400–1434.

[8]     P. Kazienko, K. Musiał, and T. Kajdanowicz, Multidimensional Social Network and Its Application to the Social Recommender System, *IEEE Transactions on Systems, Man and Cybernetics - Part A: Systems and Humans*, Vol. 41, Issue 4, 2011, pp. 746-759.

[9]     M.A. Rodriguez, A multi-relational network to support the scholarly communication process, International Journal of Public Information Systems, vol. 1, 2007, pp. 13–29. URL http://arxiv.org/abs/cs/0601121

[10]     M.A. Rodriguez, J. Shinavier, Exposing Multi-Relational Networks to Single-Relational Network Analysis Algorithms, Journal of Informetrics, 2009, 4(1), pp. 29-42.

[11]     M. Szella, R. Lambiotte, S. Thurnera, Multirelational organization of large-scale social networks in an online world, Proceedings of the National Academy of Science of the United States of America, 2010, 107(31), pp. 13636-13641.

[12]     L. Garton, C. Haythorntwaite, and B. Wellman, Studying Online Social Networks, *Journal of Computer-Mediated Communication*, vol. 3, no. 1 1997, pp. 75-105.

[13]     J. Golbeck, and J. Hendler, FilmTrust: movie recommendations using trust in web-based social networks, in *Proc. Consumer Communications and Networking Conference*, IEEE Conference Proceedings, vol. 1, 2006, pp. 282-286.

[14]     B. Wellman, J. Salaff, D. Dimitrova, L. Garton, M. Gulia, and C. Haythornthwaite, Computer Networks as Social Networks: Collaborative Work, Telework, and Virtual Community, *Annual Review of Sociology*, vol. 22, no. 1, 1996, pp. 213-238.

[15]     M. Girvan, and M.E.J. Newman, Community structure in social and biological networks, in *Proc. the National Academy of Sciences*, USA, vol. 99, no. 12, 2002, pp. 7821-7826.

[16]     N. Agarwal, M. Galan, H. Liu, and S. Subramanya, WisColl: Collective Wisdom based Blog Clustering, *Information Sciences*, vol. 180, no. 1, 2010, pp. 39–61.

[17]     P. Bródka, K. Musiał, and P. Kazienko, A Performance of Centrality Calculation in Social Networks. In *Proc. CASoN 2009 International Conference on Computational Aspects of Social Networks*, IEEE Computer Society, 2009, pp. 24-31.

[18]     V.D. Blondel, J.-L. Guillaume, R. Lambiotte, and E. Lefebvre, Fast unfolding of communities in large networks, *Journal of Statistical Mechanics*, P10008, 2008.

[19]     B. Huberman, D. Romero, and F. Wu, Social networks that matter: Twitter under the microscope. *First Monday*, 2009, pp. 1-5.

[20]     N.B. Ellison, C. Steinfield, and C. Lampe, The benefits of Facebook "friends: Social capital and college

students use of online social network sites, *Journal of Computer-Mediated Communication*, vol. 12, no. 4, 2007, article 1,

[21]    X. Cheng, C. Dale, and J. Liu, Statistics and social networking of YouTube videos, in *Proc. the 16th International Workshop on Quality of Service*, IEEE, 2008, pp. 229-238.

[22]    A. Capocci, V. Servedio, F. Colaiori, L. Buriol, D. Donato, S. Leonardi, and G. Caldarelli, Preferential attachment in the growth of social networks: The internet encyclopedia Wikipedia, *Physical Review E*, vol. 74, no. 3, id. 036116, 2006.

[23]    P. Domingos, Prospects and Challenges for Multi-Relational Data Mining, ACM SIGKDD Explorations Newsletter, vol. 5, no. 1, 2003.

http://jcmc.indiana.edu/vol12/issue4/ellison.html.

[24]    D. Cai , Z. Shao , X. He, X. Yan , J. Han, Community mining from multi-relational networks, In Proceedings of the 9th European Conference on Principles and Practice of Knowledge Discovery in Databases, LNCS 3721, Springer, 2005, pp. 445-452.

[25]    P.J. Mucha, T. Richardson, K. Macon, M.A. Porter, and J.-P. Onnela, Community Structure in Time-Dependent, Multiscale, and Multiplex Networks, Science, 2010, 328(5980), pp. 876-878.

[26]    H. Zhuge, L. Zheng, Ranking semantic-linked network, in: Proceedings of the International World Wide Web Conference, Budapest, Hungary, 2003, posters.

[27]    B. Aleman-Meza, C. Halaschek-Wiener, I. B. Arpinar, C. Ramakrishnan, A. P. Sheth, Ranking complex relationships on the semantic web, IEEE Internet Computing, vol. 9, no. 3, 2005, pp. 37-44.

[28]    P. Bródka, P. Stawiak, P. Kazienko, Shortest Path Discovery in the Multi-layered Social Network. In *Proc. ASONAM 2011, The 2011 International Conference on Advances in Social Network Analysis and Mining*, IEEE Computer Society, 2011, pp. 497-501.

[29]    S. Lin, Interesting instance discovery in multi-relational data, in: D. L. McGuinness, G. Ferguson (Eds.), Proceedings of the Conference on Innovative Applications of Artificial Intelligence, MIT Press, 2004, pp. 991–992.

[30]    C. Cantador, P. Castells, Multilayered Semantic Social Network Modeling by Ontology-Based User Profiles Clustering: *Application to Collaborative Filtering, Managing Knowledge in a World of Networks,* 2006, pp. 334-349, doi:10.1007/11891451_30.

[31]    C.H. Proctor, and C.P. Loomis, *Analysis of sociometric data, in: Research Methods in Social Relations*, in M. Jahoda, M. Deutch, S.W. Cok, Ed., Dryden Press, NewYork, 1951, pp. 561-586.

[32]    M.E. Shaw, Group structure and the behavior of individuals in small groups, *Journal of Psychology*, vol. 38, 1954, pp. 139-149.

[33]    P. Kazienko, P. Bródka, K. Musiał, and J. Gaworecki, Multi-layered Social Network Creation Based on Bibliographic Data, *The Second IEEE International Conference on Social Computing (SocialCom2010)*, August 20-22, 2010, Minneapolis, IEEE Computer Society Press, USA 2010, pp. 407-412.

[34]    P. Bródka, K. Musiał, and P. Kazienko, A Method for Group Extraction in Complex Social Networks, in *Proc. WSKS 2010, The 3rd World Summit on the Knowledge Society*, Communications in Computer and Information Science, CCIS 111, Springer, 2010, pp. 238-247.

[35]    P. Bródka, K. Skibicki, P. Kazienko, and K. Musiał, A Degree Centrality in Multi-layered Social Network. In *Proc. CASoN 2011, The International Conference on Computational Aspects of Social Networks*, 2011, IEEE Computer Society, 2011, pp. 237-242.

[36]    N. Ferreira, Social Networks and Young People: A Case Study. International Journal of Human Capital and Information Technology Professionals (IJHCITP), vol. 1, no. 4, 2010, pp. 31-54.

[37]    A. García-Crespo, R. Colomo-Palacios, J.M. Gómez-Berbís, and B. Ruiz-Mezcua, SEMO: a framework for customer social networks analysis based on semantics. Journal of Information Technology, vol. 25, no. 2, 2010, pp. 178-188.

[38]    A.J. Berlanga, F.J. García Peñalvo, and P.B. Sloep, Towards eLearning 2.0 University. Interactive Learning Environments, 2010, vol. 18, no. 3, pp. 199-201.

[39]    P. Kazienko, and T. Kajdanowicz, Label-dependent Node Classification in the Network. *Neurocomputing*, Vol. 75, Issue 1, 2012, pp. 199-209.